\documentclass[journal,twoside,web]{IEEEtran}

\usepackage{cite}
\usepackage{amsmath,amssymb,amsfonts}
\usepackage{algorithmic}
\usepackage{graphicx}
\usepackage{textcomp}
\usepackage{placeins}
\usepackage{float}
\usepackage{hyperref}
\usepackage{multirow}
\usepackage{multicol}

\def\BibTeX{{\rm B\kern-.05em{\sc i\kern-.025em b}\kern-.08em
    T\kern-.1667em\lower.7ex\hbox{E}\kern-.125emX}}

\begin{document}

\title{Conditioning Deep Anatomical Prior Knowledge for Reconstruction of Multispectral Optoacoustic Tomography Images}
\author{Sarah Franceschin, Lukas Imanuel Scheel-Platz, Philipp Haim, Guillaume Zahnd, Vasilis Ntziachristos, and Dominik J\"ustel
\thanks{This project has received funding from the European Research Council (ERC) under the European Union’s Horizon Europe research and innovation programme under grant agreement No 101041936 (EchoLux) (Corresponding Author: Dominik
J\"ustel, e-mail: dominik.juestel@helmholtz-munich.de) }
\thanks{Sarah Franceschin, Lukas Imanuel Scheel-Platz, Philipp Haim, Vasilis Ntziachristos, and Dominik J\"ustel are with the Chair of Biological Imaging, Central Institute for Translational Cancer Research (TranslaTUM), School of Medicine and Health \& School of Computation, Information and Technology, Technical University of Munich, 81675 Munich, Germany, and also with the  Institute of Biological and Medical Imaging, Bioengineering Center, Helmholtz Zentrum München, 85764 Neuherberg, Germany}
\thanks{Sarah Franceschin, Lukas Imanuel Scheel-Platz, Philipp Haim and Dominik J\"ustel are also with the Institute of Computational Biology, Helmholtz Zentrum München, 85765 Neuherberg, Germany, and also with the Institute of AI for Health, Helmholtz Zentrum München, 85764 Neuherberg, Germany}
\thanks{Lukas Imuanel Scheel-Platz and Philipp Haim are also with the Ludwig-Maximilians-Universit\"at
M\"unchen, 80539 Munich, Germany.}
\thanks{Guillaume Zahnd was with iThera Medical GmbH, Munich, Germany.}}

\IEEEpubid{This work has been submitted to the IEEE for possible publication. Copyright may be transferred without notice, after which this version may no longer be accessible.}
\maketitle

\begin{abstract}
Accurately delineating tissues and reconstructing their chromophore compositions from Multispectral Optoacoustic Tomography (MSOT) images is a key challenge in optoacoustic imaging. The difficulty arises because light fluence distributions within tissue intrinsically depend on spectral optical properties, making the inverse problem inherently ill-posed. Currently, there is a lack of studies leveraging a priori probabilistic anatomical knowledge to guide tissue segmentation and infer chromophore composition. Moreover, most current studies address these two tasks sequentially, which can result in errors accumulating through the process.
To address these issues, we present Anatomical Priors for Reconstruction of Optoacoustic Tomography (APRECOT), a method that leverages probabilistic models of anatomical structures and tissue properties, to enable simultaneous segmentation of tissues and reconstruction of their bulk chromophore compositions. In this proof-of-concept using \textit{in-silico} data, we show that incorporating probabilistic anatomical context strongly improves the accuracy of bulk chromophore concentration estimation compared to reference methods that do not use any anatomical context or use sequential strategies. This work represents an essential step towards an MSOT imaging mode that directly provides clinically relevant information, such as imaging tissue oxygenation dynamics or disease-related changes in tissue composition. 

\end{abstract}
\IEEEpubidadjcol 

\section{Introduction}
\label{sec:introduction}
\enlargethispage{-1.5cm}
Spectral analysis performed on optoacoustic images obtained at multiple wavelengths is commonly applied to separate and quantify the underlying tissue chromophores and yield insights into tissue composition, including levels of deoxy- and oxy- hemoglobin (HHb, HbO$_2$), lipids, melanin, as well as tissue function such as oxygen saturation (SO$_2$) \cite{ Roll2019, Karlas2020, Karlas2021, Jstel2023, Ntziachristos2025}. By focusing the analysis on specific regions of interest (ROIs), such as blood vessels or muscles, it is possible to investigate localized imaging features that relate directly to a biological question or the disease being investigated. However, accurate delineation of ROIs in MSOT images and estimation of their chromophore concentrations can be challenging, as wavelength-dependent light attenuation in tissue makes the problem ill-posed and leads to ambiguous results \cite{Ntziachristos2010}. Previous work demonstrated that exploiting anatomical context can help mitigate the ill-posedness of the reconstruction problem and improve spectral analysis performance by regularizing the solution space according to expected tissue organization  \cite{Mandal2016, Liang2022, Pattyn2021, Brochu2017, Zhang2022, Grasso2022}. However, if the anatomical context consists of rigid, deterministic constraints, it can introduce systematic bias into the reconstruction and compromise chromophore quantification, as inter-subject anatomical variability, imperfect image registration, or user-dependent segmentation errors are not taken into account. Moreover, when ROI identification and spectral analysis are treated as two independent, sequential steps, opportunities for intercommunication are lost, which increases susceptibility to error accumulation.

In optoacoustic (OptA) imaging, most existing methods leveraging  anatomical context can be categorized in two groups: segmentation-based direct correction (SBDC) methods \cite{Mandal2016, Liang2022, Pattyn2021} and segmentation-based iterative correction (SBIC) methods \cite{Brochu2017, Zhang2022, Grasso2022}. These methods typically rely on fixed anatomical knowledge derived from manual or automated segmentation of co-registered ultrasound (US) or Magnetic Resonance (MR) images. Such priors are applied directly for fluence correction (SBDC), or serve as initial inputs within algorithms that iteratively refine the estimation of tissue optical properties (SBIC). Another study has instead leveraged US-derived anatomy to directly inform the model-based acoustic reconstruction with intra-tissue homogeneity assumptions \cite{Yang2020}. However, all these methods are prone to increased error because they perform tissue segmentation and spectral unmixing as independent, sequential steps.  Moreover, none of these methods exploits prior knowledge of tissue locations and optical properties in the form of probability distributions, which is important for generalizability. Integration of probabilistic anatomical models has been previously demonstrated only for reconstruction tasks of different imaging modalities, such as MR \cite{Tohka2014}. 

A previous study has demonstrated that knowledge on tissue spatial locations can be represented probabilistically by means of generative models \cite{Schellenberg2022b}. In a follow-up study \cite{Nlke2024}, the same probability distribution was used to train an invertible neural network predicting oxygen saturation posterior distributions. Here, each anatomical structure was associated to specific distributions of optical absorption properties, defined by mean and standard deviation values obtained from literature. However, this method is limited in estimating oxygen saturation within blood vessels. Moreover, the framework was only used with pixel-wise inputs, thereby failing to exploit image-level information that provides spatial context.

Coupling image segmentation and functional reconstruction has been previously demonstrated successfull by other imaging modalities such as Positron Emission Tomography (PET) \cite{Bagci2013, Storath2015}.
In OptA imaging, only a few coupled approaches have been explored, primarily focusing on combining segmentation with acoustic reconstruction or with oxygen saturation inference \cite{Luke2019, Boink2020, Shang2024}. However, none of these approaches directly provide information about tissue chromophore contents. 

In other words, existing approaches either do not leverage anatomical context in the form of probability distributions, or only deal with pixel-wise inputs.
Moreover, no current method enables automatic and concurrent tissue segmentation and determination of tissue-specific chromophore composition.

Since tissue morphology and chromophore composition are directly linked, knowledge of the underlying tissue organization can inform plausible chromophore distributions, while spectral information may itself provide cues regarding tissue identity. We therefore hypothesize that leveraging anatomical context as joint probability distributions of tissue localization and tissue-specific properties, while enabling intercommunication between segmentation and chromophore mixture predictions, can improve the accuracy of spectral analysis within ROIs. Therefore, in this work we present a method that uses anatomical prior knowledge, namely probability distributions of the spatial organization of anatomical structures and tissue-specific chromophore compositions, to simultaneously segment multiple tissue types and extract their bulk chromophore concentrations (i.e. best-fit mixture for tissue type) from \textit{in-silico} MSOT images. By comparing our method, APRECOT, with two distinct reference methods, we demonstrate that using anatomical prior knowledge rather than relying on spatially homogeneous priors, and coupling tissue segmentation and chromophores unmixing instead of performing them sequentially improve the chromophore mixture inference by 77\% and 88\%, respectively.

\section{Methods}
\begin{figure*}
    \centering
    \includegraphics[width=\textwidth]{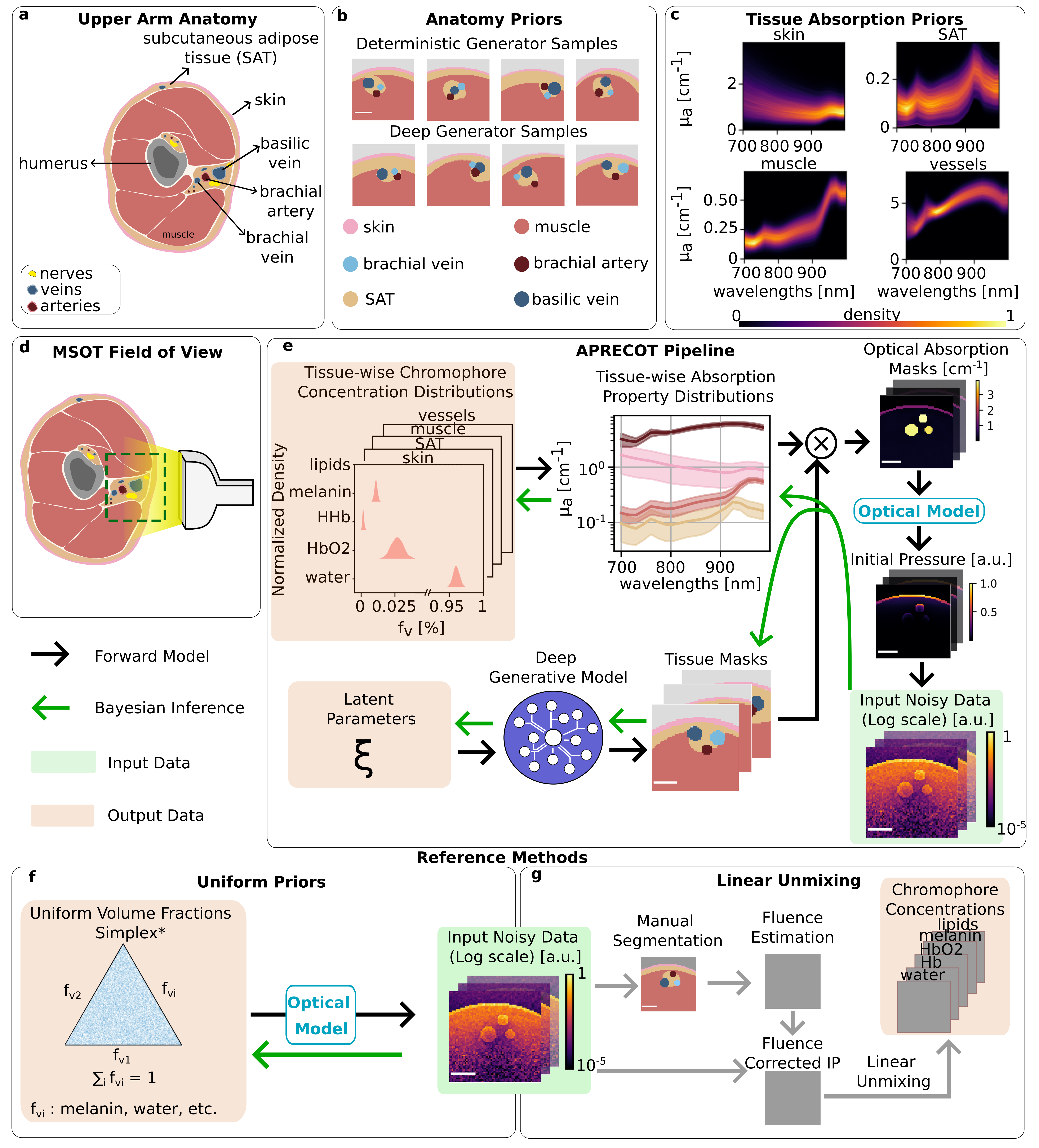}
    \caption{Schematic overview of the workflow for our Anatomical Priors for Reconstruction of Optoacoustic Tomography (APRECOT) method. [a] Cross-sectional schematic of a computer-simulated human upper arm, which is the region of interest in this study. [b-c] Priors derived from anatomical knowledge. [b] Prior knowledge on anatomy encoded as tissue masks, which delineate the main anatomical structures (skin, SAT, muscle, brachial vein, basilic vein, and brachial artery). [c] Absorption spectrum ($\mu_a$) priors for the four modeled tissue types (i.e., skin, SAT, muscle, vessels). [d] The in-silico images used for this study approximate the area of interest outlined by the dashed rectangle, assuming the MSOT probe is positioned according to the figure. [e] APRECOT method workflow. The tissue composition distributions define the tissue absorption spectra distributions. Tissue masks are generated by a deep generative model that uses latent parameters as input. Each tissue absorption mask is then defined by assigning the appropriate tissue absorption spectrum to the corresponding tissue mask. Subsequently, an optical model transforms these tissue absorption masks into initial pressure images. The APRECOT method receives noisy initial pressure data as input and applies Variational Bayesian inference to invert the forward model, thereby inferring tissue composition distributions and the latent parameters of the deep generative model. [f-g] Reference methods. [f] A reference method that uses uniform prior knowledge. Here, a 2-simplex (3 total volume fractions) instead of a 4-simplex (5 total volume fractions) is shown for simplicity. The same optical model as APRECOT is used to obtain initial pressure images from chromophore distributions. Variational Bayesian inference estimates chromophore concentrations from the input data. [g] Sequential reference method: (1) manual image segmentation, (2) assigning average absorption properties to the different tissues and simulating fluence, (3) fluence correction, (4) linear unmixing. All the scalebars indicate 1 cm. MSOT - multispectral optoacoustic tomography, HHb - deoxyhemoglobin, HbO$_2$ – oxyhemoglobin, f$_v$ – volume fraction, IP - initial pressure, a.u. - arbitrary units, SAT – subcutaneous adipose tissue, log – logarithmic.}
    \label{fig:methods}
\end{figure*}

\subsection{An ill-posed inverse problem}

    In OptA imaging, short laser pulses are absorbed by endogenous chromophores, leading to a rapid thermoelastic expansion that generates an initial pressure distribution $p_0$ \cite{Ntziachristos2010}. The initial pressure propagates as acoustic waves, finally captured by the detectors. The entire OptA process can be described by concatenating the models defining light absorption and sound propagation in tissue. Assuming the Grüneisen coefficient constant and known, the light absorption can be modeled as
	\begin{equation}
    \label{eq:opticalmodel}
        p_0 \simeq \mu_a\Phi(\mu_a, \mu_s),
    \end{equation} where $\Phi$ describes the fluence, and $(\mu_a$, $\mu_s)$ are the absorption and scattering properties of the tissue, respectively \cite{RevModPhys.97.015005}. The latter are related to the chromophore volume fractions $f_{v,i}$ in the imaged tissue according to:
    \begin{equation}
		\label{eq:tissueproperties}
		\mu_a = \sum_{i} f_{v,i}\mu_{a,i}
	\end{equation} with $\mu_{a,i}$ being the pure absorption spectra of chromophore $i$ \cite{Jacques2013}. The optical model is strongly non-linear, as the fluence $\Phi$ depends implicitly on the absorption and scattering coefficients $\mu_a$ and $\mu_s$. Thus, retrieving the quantity of interest $f_{v,i}$ from $p_0$ (i.e. spectral unmixing), requires to solve an ill-posed inverse problem. 

\subsection{Bayesian formulation}
    Adopting the conditioning strategy developed in \cite{Knollmller2021}, this proof of concept aims to reconstruct joint tissue segmentations $seg$ and best-fit chromophore mixtures $bcm$ from initial pressure images $p_0$ in the context of the Bayes’ theorem:
    \begin{equation}
        P(seg, bcm\space|\space p_0) \propto P(p_0\space|\space seg,\space bcm)P(seg, bcm),
        \label{eq:bayes}
    \end{equation}
    
    The likelihood $P(p_0\space|\space seg, bcm)$ was modeled using the forward optical model (Eq. \ref{eq:opticalmodel}). In particular, we leveraged \textit{FiPy} \cite{Guyer2009}, a finite volume PDE solver for Python, to implement the diffusion approximation (DA) \cite{Martelli2022} of light transport in scattering tissue so that it is differentiable, and thus conditionable. Finally, the anatomical prior $P(seg, bcm)$ was modelled with probability distributions of the spatial organization of anatomical structures and tissue-specific properties. $P(seg, bcm)$ is described in the following section. 

 \subsection{Anatomical Prior Knowledge}
 \label{subsec:anatpriors}
\subsubsection{Anatomy}
\label{subsubsec:anatomy}
    This work models the main anatomical structures of the upper-arm, with a particular focus on the medial bicipital groove (MBG) (Fig. \ref{fig:methods}a). To encode anatomical prior knowledge, an anatomical generator was utilized. Initially, we developed a stochastic generator capable of simulating upper arm anatomy in two dimensions (StAG – Stochastic upper Arm Generator). To ensure differentiability and increase generalization, a deep generative model (DeepAG – Deep upper Arm Generator) was trained using a training dataset generated by StAG.
    
    StAG generates segmentation maps of cross-sections of the upper arm, crops 4$\times$4 cm$^2$ regions that contain the MBG, and discretizes the cropped images to 64x64 pixels. StAG works by encoding anatomical structures as geometrical shapes (such as circles, ellipses, and rings), with size and position originally sampled from specific probability distributions, adjusted or conditionally accepted to satisfy anatomical constraints (code available at \url{https://github.com/juestellab/anatomical-generator}).  
    
    For training DeepAG, we compiled a training dataset of 500,000 images from StAG (Fig. \ref{fig:methods}b). We ensured gradual variations in anatomical structures between the samples by generating sequences of images where some structures remained fixed while others varied. 
    For DeepAG we used a consistency model (CM) due to its computational efficiency \cite{Song2023}, with  a hierarchical U-Net as underlying architecture \cite{Ronneberger2015, kinyugo2025consistency_models}. Using a CM requires choosing a set of noise levels for the noising and denoising steps. In our case, $\sigma$ = [80.0, 1.2, 0.661] was empirically determined as a good compromise between speed and image quality. \textit{Pytorch} and \textit{Pytorch Lightning} \cite{Falcon2019} were used to implement the network and the training strategy. We trained the CM with L2-loss over approximately 36 hours on an NVIDIA A100 GPU. We validated the expressiveness of the trained model via UMAP embeddings \cite{McInnes2018} of the generated data (see Fig. \ref{fig:umaps} in the Appendix).

\subsubsection{Tissue-specific Properties}
	\label{subsubsec:tissueproperties}
    To model the tissue-specific properties (Fig. \ref{fig:methods}c), we leveraged the linear mixture model expressed in Eq. \ref{eq:tissueproperties}. Hb, HbO$_2$, H$_2$O, lipids, melanin were considered together with a generic, non-absorbing component, such that $\sum_i f_{v, i} = 1$. The corresponding pure spectra $\mu_i$ were found in literature \cite{Prahl2018, Segelstein1981, Altshuler2003, Jacques2018}. $f_{v, i}$ were obtained with $f_{v, i} = \frac{s(X_i)}{\sum_j s(X_j)}$ , where $X_i \sim \mathcal{N}(\mu_i, \sigma_i)$, and $\space s:\mathbb{R}\rightarrow[v_1, v_2]$ is the soft-clipping function. The bounds $v_1, v_2$ reflect physiologically meaningful limits. 
    $\mu_i$ and $\sigma_i$ were retrieved after conducting an extensive literature review to understand the chromophore compositions of the tissues of interest: skin, subcutaneous adipose tissue (SAT), muscle, vessels (Table \ref{table:volfrac}).

\begin{table}
\centering
\setlength{\tabcolsep}{2pt}
\renewcommand{\arraystretch}{1.2}
\begin{tabular}{|c|cc|c|c|}
\hline
\multirow{2}{*}{\textbf{Tissue}}  & \multirow{2}{*}{\textbf{Chromophore}} & $\boldsymbol{F_{v,i}}$ & \textbf{SO$_2$} & \multirow{2}{*}{\textbf{Ref.}} \\
 & & Mean (Std) & Mean (Std) & \\
\hline
\multirow{2}{*}{epidermis} & melanin & 8.7\% (3.7\%) &\multirow{2}{*}{-} & \multirow{4}{*}{\parbox[c]{1.2cm}{\centering \cite{Jacques2013, Bashkatov2011, Jacques1998, Mizukoshi2022}}} \\ 
     & water &  65\% (5\%) &  &  \\ \cline{1-4}

  \multirow{2}{*}{dermis} & water & 70.2\% (5\%) &\multirow{2}{*}{59\% (3.5\%)} & \\ 
    & blood & 2\% (1\%)&  &  \\ 

\hline
\multirow{3}{*}{\parbox[c]{1.2cm}{\centering subcut. adipose tissue}} & lipids & 80\% (10\%) & \multirow{3}{*}{60\% (10\%)} & \multirow{3}{*}{\parbox[c]{1.2cm}{\centering \cite{Bashkatov2011, AnnalsoftheICRP1967, nist_adipose_tissue, Mizukoshi2022}}}\\ 
& water & 15\% (5\%)&   & \\ 

    & blood & 2\% (1\%)&  & \\
\hline

\multirow{3}{*}{muscle} & water & 72\% (5\%) & \multirow{3}{*}{73\% (2.5\%)} & \multirow{3}{*}{\parbox[c]{1.2cm}{\centering \cite{Bashkatov2011, AnnalsoftheICRP1967, Ward2005, Miranda-Fuentes2021}}}\\ 
& blood & 3\% (1\%) &  &\\
& lipids & 2.5\% (0.3\%)&  &\\
\hline

\multirow{3}{*}{vessel} & cells (RBCs) & 45\% (3\%)&  \multirow{3}{*}{82\% (16\%)} &  \multirow{3}{*}{\parbox[c]{1.2cm}{\centering \cite{Guyton2021, Roggan1999, Nitzan2008}}}\\ 
 & Hb in RBCs &  328$[\frac{g}{L}]$ (14.5$[\frac{g}{L}]$) & &\\ 
& plasma (water) & $\neg$cells & & \\
\hline
\end{tabular}
	 \caption{Mean and standard deviation (Std) values for the tissue-wise chromophore volume fractions. rbcs - Red Blood Cells.}
	 \label{table:volfrac}
\end{table}

\subsection{The Reconstruction}
\label{subsec:anatrec}
 To estimate the posterior $P(seg, bcm\space|\space p_0)$ (Eq. \ref{eq:bayes}) from initial pressure images (Fig. \ref{fig:methods}d-e), we leveraged NIFTy \cite{Edenhofer2024}, a package implementing Information Field Theory (IFT) methods. Specifically, Metric Gaussian Variational Inference (MGVI) \cite{Knollmller2019}, a Bayesian variational inference method, was used. By using MGVI, it was possible to condition the anatomical prior knowledge with initial pressure images to find a set of plausible samples from the approximate joint posterior distribution of segmentations and bulk chromophore compositions, capturing their correlation structure as well. The code is available at \url{https://github.com/juestellab/APRECOT}.
 \label{sec:methods}
\begin{figure*}
    \centering
    \includegraphics[width=0.91\textwidth]{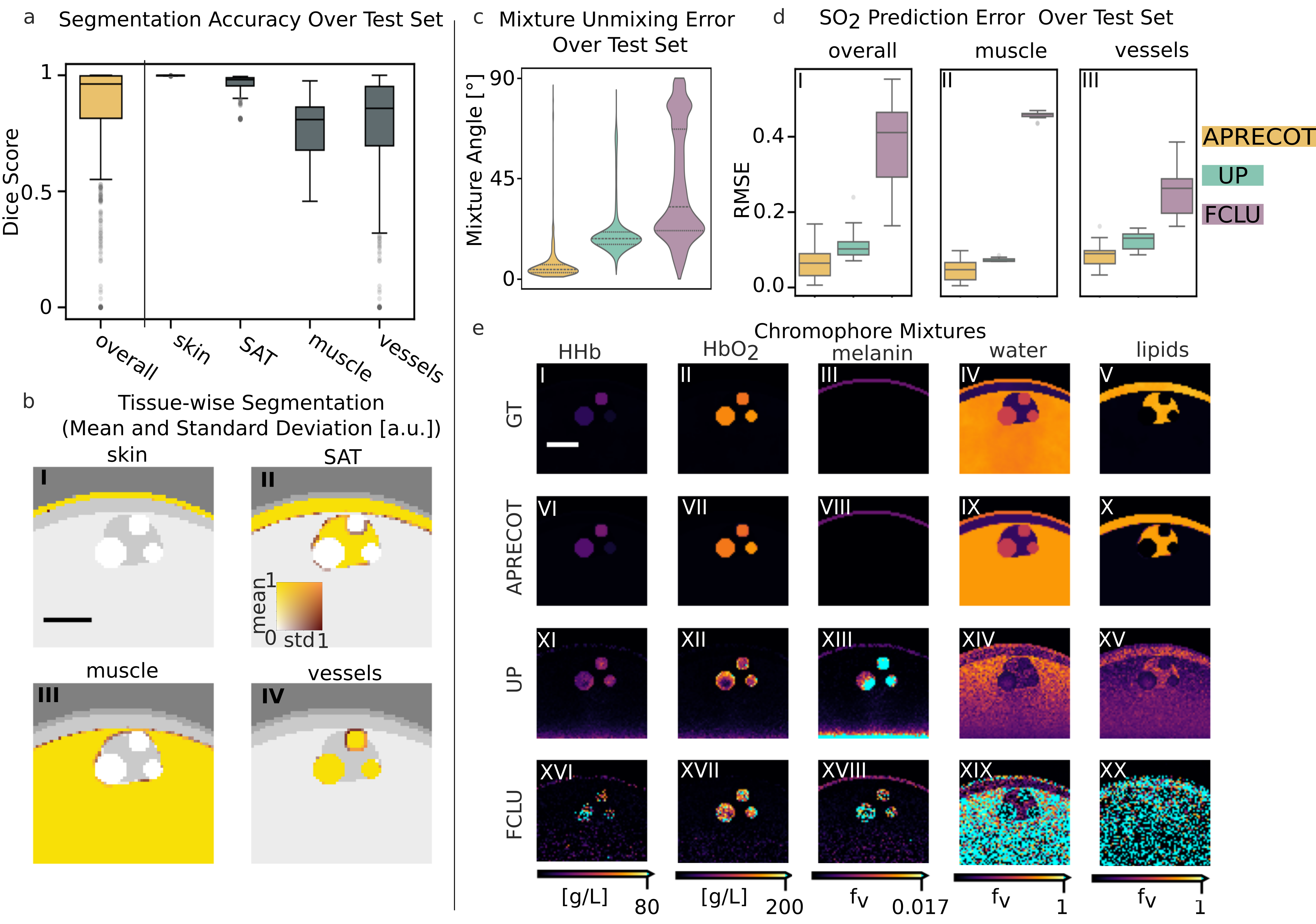}
    \caption{Segmentation, as well as qualitative and quantitative results for chromophore unmixing. [a] Distribution of Dice Scores computed using segmentation predictions from the full test set for the whole image and for each tissue. [b] Tissue-wise segmentation for a representative test set sample, with the ground truth anatomical layout shown in grayscale. The segmented region’s mean and standard deviation is overlayed using a bivariate colormap. [c] Distributions of mixture angles computed with Anatomical Priors for Reconstruction of Optoacoustic Tomography (APRECOT), Uniform Priors (UP) and Fluence Corrected Linear Unmixing (FCLU), using the full test set. [d] Oxygen saturation (SO$_2$) prediction error distributions of the whole image, muscle and vessels, computed with APRECOT, UP and FCLU, using the full test set. [e] Chromophore concentrations of the representative sample depicted in [b] are shown for ground truth (GT) and those inferred using APRECOT, UP and FCLU. SAT - subcutaneous adipose tissue, HHb - deoxyhemoglobin, HbO$_2$ - Oxyhemoglobin, f$v$ - volume fraction, RMSE - Root mean squared error. All scalebars indicate 1 cm.}
    \label{fig:results1}
\end{figure*}

    \subsection{Validation}
    \label{subsec:Validation}
    We validated our method using a test set of 12 \textit{in-silico} initial pressure images. The anatomical structures were sampled from StAG, ensuring no duplicated data was used for training DeepAG. For each image, tissue-specific chromophore compositions, and therefore absorption properties, were randomly sampled from the probabilistic model defined in \ref{subsubsec:tissueproperties}. Moreover, we simulated three levels of spatial inhomogeneity (0\%, 50\%, and 100\% of the amplitude range of the prior) in the absorption properties for SAT and muscle regions. We used 2D Gaussian processes as defined in \cite{Arras2022} to describe the spatial correlations. For each inhomogeneithy level, our method was benchmarked against two reference methods (Fig. \ref{fig:methods}f-g), to evaluate: (1) the advantage of using anatomical context over spatially homogeneous priors, and (2) the advantage of performing segmentation and chromophore mixture inference simultaneously rather than sequentially.

    For the first reference method, we used the same variational inference framework (NIFTy) and optical model applied in APRECOT. Here, prior knowledge considers the vector of volume fractions at each spatial location as independent draw from a uniform distribution on the standard simplex. No segmentation was involved. We refer to this reference method as Uniform Priors (UP) (Fig. \ref{fig:methods}f). For the second reference method, we developed a SBDC method inspired by \cite{Mandal2016, Pattyn2021, Liang2022}, which exploited segmentations to correct fluence decay (Fig. \ref{fig:methods}g). We assumed perfect segmentation in order to focus on the tissue composition inference. We assigned average absorption properties $\overline{\mu_a}$ to each of the tissues (derived from the probabilistic model in \ref{subsubsec:tissueproperties}) and simulated fluence using the optical model described in \ref{subsec:anatrec}. The absorption properties were estimated by performing fluence correction:
    \begin{equation}
        \mu_a = \frac{p_0}{\Phi(\overline{\mu_a})},
    \end{equation}
    where $\Phi(\overline{\mu_a})$ is the simulated fluence. Finally, linear unmixing is performed to estimate the chromophore concentrations. We refer to this method as Fluence Corrected Linear Unmixing (FCLU). 
    Performance was assessed by evaluating tissue segmentation accuracy and chromophore composition fidelity, using the Dice Score (DS) and the Spectral Angle (SA) \cite{Kruse1993}, respectively.
    The SA measures the similarity between two spectral vectors and is defined as:
    \begin{equation}
    \alpha = \arccos \left( \frac{s_1 \cdot s_2}{\lVert s_1 \rVert \, \lVert s_2 \rVert} \right) \in [0, \pi/2],
    \end{equation}
    where $s_1$ and $s_2$ are the reference and predicted spectra, respectively. A lower value of $\alpha$ indicates higher spectral similarity.
    Chromophore mixtures are represented as spectra with dimensions $(N_x, N_y, N_{chr})$, where $N_x$ and $N_y$ denote the spatial dimensions and $N_{chr}$ the number of chromophores.
    Since only APRECOT produces segmentation outputs, DS was exclusively reported for this method and evaluated across all posterior samples ($n = 28$).
    To evaluate the accuracy of chromophore composition inference, pixel-wise SA was computed for all methods, resulting in spectral maps (SMs). For APRECOT and UP, SMs were computed for all posterior samples, and the mean was used for comparison. Finally, we compute the oxygen saturation SO$_2$ from the predicted chromophore mixtures as follows: 
    \begin{equation}
    SO_2 = \frac{[HbO_2]}{[HHb] + [HbO_2]},
    \end{equation} where [HbO$_2$] and [HHb] refers to the concentrations of HbO$_2$ and HHb, respectively. We compared the accuracy of SO$_2$ estimation between APRECOT, UP and FCLU using the Root Mean Squared Error (RMSE). For APRECOT and UP, RMSEs were calculated based on the mean of posterior chromophore mixture samples. Both the SMs and the RMSEs were evaluated within tissue regions, excluding background regions. The reconstructions of tissue segmentations and inference of best fit chromophore compositions for the test set required about 16 days on an NVIDIA A100 GPU.

\section{Results}
\label{sec:results}
As described in Sec. \ref{subsec:Validation}, APRECOT, UP and FCLU were benchmarked against three datasets having different spatial inhomogeneity levels in the absorption properties. The results obtained with Level 2 inhomogeneities (50\% of the priors amplitude) are described in this section and shown in Fig. \ref{fig:results1}, while the results for the others are summarized in Tables \ref{table:segperf}-\ref{table:inhomoglev} of the Appendix.
APRECOT demonstrated good segmentation accuracy, with a median DS of 0.96 (Fig. \ref{fig:results1}a). The boxplots in Fig. \ref{fig:results1}a reveals that skin and SAT were segmented with higher accuracy (median of 0.99 and 0.98 respectively), whereas muscle and vessels were harder to accurately segment, with lower median DS of 0.81 and 0.86, respectively. We show an example of segmentation prediction in Fig. \ref{fig:results1}b, where the bivariate colormap illustrates the ratio of the mean to the standard deviation from the posterior tissue segmentations samples: predictions with higher certainty appear more yellow, while increased uncertainty is indicated by darker red tones. As expected, the inter-tissue boundaries are the main source of uncertainty. 

Our results show that APRECOT achieves higher accuracy in chromophore mixture inference compared to UP and FCLU, evidenced by median mixture angle reduction of 77\% and 88\%, respectively Fig. \ref{fig:results1}c. Moreover, APRECOT outperfromed UP and FCLU in estimating SO$_2$, reducing the median RMSE by 38\% and 85\%, respectively (Fig. \ref{fig:results1}d-I). The RMSE reduction was also computed for muscle and vessels, which are ROIs where SO$_2$ is typically relevant. APRECOT reduced the median RMSE by 34\% and 89\% for muscle, and 28\% and 67\% for vessels, when compared to UP and FCLU, respectively (Fig. \ref{fig:results1}d-II,III).

Fig. \ref{fig:results1}e unveils the advantage of using anatomical prior knowledge by comparing the ground truth and predicted chromophore mixtures using APRECOT, UP and FCLU for a representative sample of the test set. In particular, the melanin channel (Fig. \ref{fig:results1}e-III, VIII, XIII, XVIII) demonstrates how only APRECOT correctly predicts no melanin in vessels, which aligns with existing literature.

APRECOT demonstrated good accuracy for bulk reconstruction of chromophore mixtures even in presence of spatial inhomogeneity. This result is qualitatevely shown in the water channel in Fig. \ref{fig:results1}e-IV, IX, XIV, XIX. While in the ground truth the water is visibly inhomogeneously distributed within the muscle (Fig. \ref{fig:results1}e-IV), APRECOT predicts a best-fit bulk composition (Fig. \ref{fig:results1}e-IX). The robustness to inhomogeneity is confirmed by the results obtained for the three inhomogeneity levels (Table \ref{table:segperf}-\ref{table:inhomoglev} in the Appendix).

\section{Discussion}
\label{sec:discussion}
For the first time, our work demonstrates that anatomical prior knowledge can be incorporated and conditioned with in-silico OptA images, enabling concurrent segmentation of multiple tissues and reconstruction of bulk chromophore concentrations. This approach, APRECOT, is not restricted to vessels\cite{Luke2019, Boink2020, Shang2024} or pixel-wise oxygen \cite{Nlke2024} saturation inference. Moreover, this method is the first to define anatomical prior knowledge as probability distributions for tissue segmentation and chromophore composition determination and to show how to condition it by means of Bayesian Variational Inference. Our results demonstrate that by incorporating probabilistic anatomical context and exploiting synergies between segmentation and chromophore unmixing, bulk chromophore reconstruction is improved compared to methods that omit anatomical context or perform these tasks sequentially.

The exploration of synergies between tasks has become an increasingly prominent topic in medical imaging processing \cite{Boink2020, Shang2024, Luke2019,Corona2019,Kofler2024,Karkalousos2023,Sun2018}, where downstream tasks such as segmentation are commonly integrated with image reconstruction. Multitask learning (MTL) is a widely used approach, enabling a single machine-learning model to learn multiple tasks concurrently. While this work shares similarities to MTL, it differs by employing a probabilistic approach that incorporates structured priors and likelihoods. 
Our approach, which exploits task synergies, offers clear advantages. These advantages include prevention of error accumulation and reduction of bias associated with sequential processing steps. However, our current work still remains a proof of concept based solely on simulated data. Nevertheless, the modular structure of the APRECOT model makes it easy to add an acoustic model, which would enable adaptation of the approach to real-world data.

Foundation models are AI models trained on large datasets to perform a wide range of tasks, and they are emerging as the leading approach in machine learning. Their use is growing in medical imaging for tasks including segmentation, data augmentation, feature extraction, and image reconstruction \cite{Azad2023, Noh2025}. Previous studies have explored generative foundation models as priors for image reconstruction across various modalities \cite{Song2022, Dey2024, Hashimoto2025}. Our work advances efforts to integrate foundation models into OptA image reconstruction and notably employs Bayesian variational inference in this field for the first time. Including anatomical knowledge through deep generative models enables the integration of anatomical foundation models trained on other imaging modalities including MRI and CT into the reconstruction process. This approach facilitates broader population representation and enhances the generalizability of the method. Therefore, leveraging anatomical foundation models would allow encoding of more accurate anatomical structures, enabling the application of APRECOT to real OptA imaging data.

Accurate delineation of ROIs and estimation of their chromophore composition are critical for most clinical applications of OptA imaging \cite{Roll2019, Karlas2020, Karlas2021}. By simultaneously addressing both tasks, our method directly meets these clinical needs. As a result, our work is a major step towards interpretability and usability of MSOT images in clinical scenarios, which is essential for clinical translation of the technology. Another necessary step is enablilng APRECOT to work in real-time, and recent progress in deep learning strategies \cite{Dehner2023} suggest its feasibility. Deep learning models could learn the Bayesian inference operator realized by APRECOT, thereby increasing reconstruction speed and enabling real-time performance.

In summary, we present an OptA reconstruction method that leverages anatomical and tissue-specific a priori knowledge. While currently limitated to offline, in-silico data reconstruction, our method represents an important step towards developing a novel imaging modality capable of providing real-time feedback, like tissue segmentation and oxygen saturation assessment.

\section{Acknowledgment}
The authors would like to thank Dr. Suhanyaa Nitkunanantharajah, Sarkis Ter Martirosyan,  Maximilian Bader, David Gorbunov, Maria Begona Rojas Lopez, Manuel Gehmeyr and Dr. Thi Bich Tram Do for the fruitful discussions, as well as Serene Lee for her assistance in the writing process. 

\section{Conflict of Interest}
V.N. is a founder and equity owner of Maurus OY, sThesis GmbH, Spear UG, Biosense Innovations P.C. and I3 Inc. G.Z. was an employee of iThera Medical GmbH during the preparation of this manuscript.

\bstctlcite{IEEEexample:BSTcontrol}
\bibliography{library.bib}

\section{Appendix}

\label{sec:appendix}
\begin{figure}[H]
    \centering
    \includegraphics[width=0.42\textwidth]{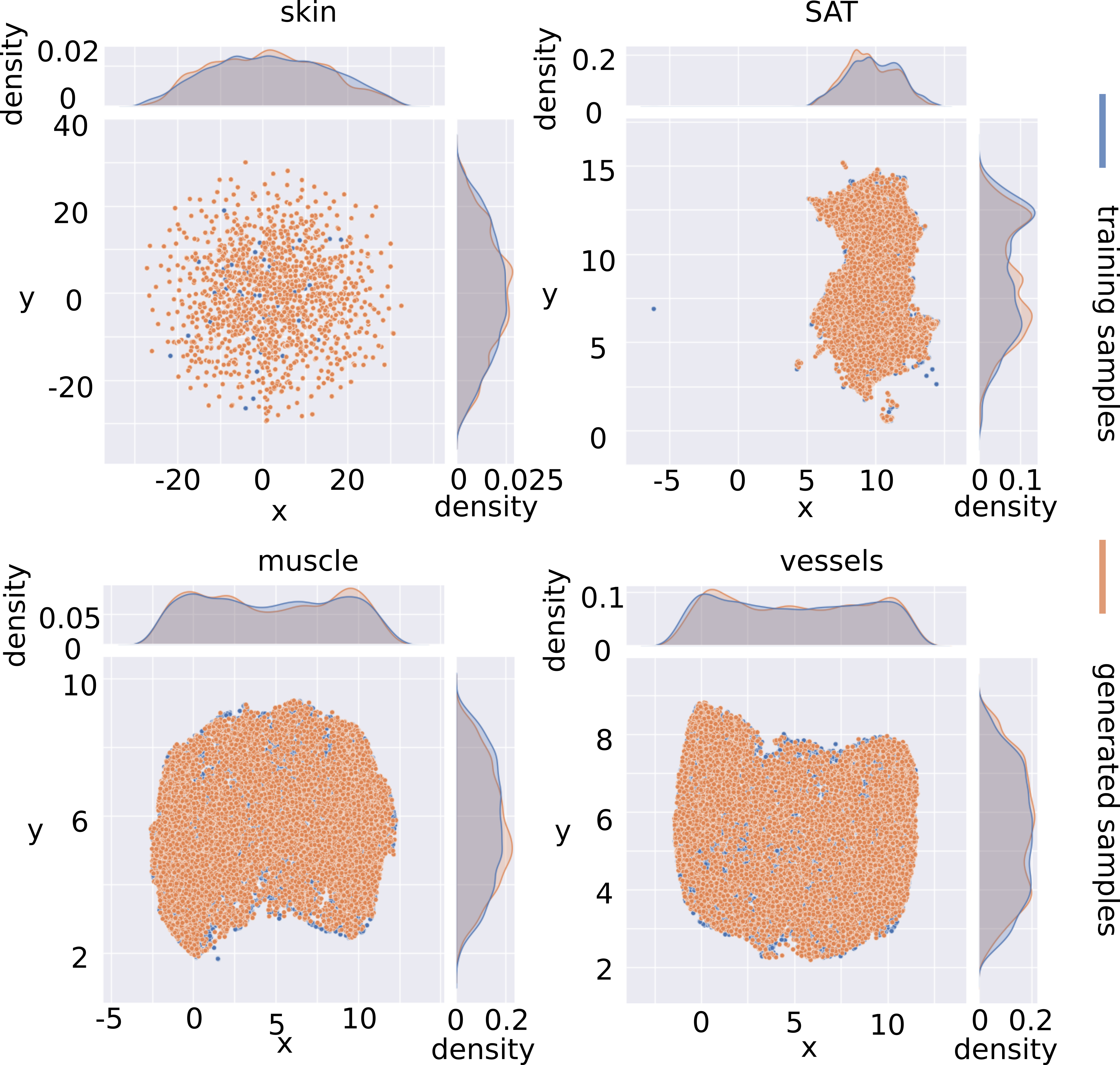}
    \caption{2D-UMAP scatter plot for each channel, where each channel represents a specific tissue. Each plot contains 10,000 training samples overlaid with generated samples. Density profiles are shown for both dimensions of the UMAP. SAT - Subcutaneous Adipose Tissue.}
    \label{fig:umaps}
\end{figure}

\begin{table}[H]
	 	\centering
	 \begin{tabular}{ |c|c|}
	 	
	 	\hline
	 	\textbf{Inhomogeneity} & \textbf{Median DS for segmentation} \\
         & \textbf{and their ranges (in brackets)} \\
	 	\hline
	 	Level 1 & 0.968 (0.861 – 0.998)\\
	 	\hline
        Level 2 & 0.964 (0.816 – 0.997)\\
	 	\hline
        Level 3 & 0.968 (0.840 – 0.997)\\
        \hline
	 \end{tabular}
	 \caption{Segmentation performance of APRECOT for three levels of inhomogeneity, expressed as median dice score (DS) and interquartile range.}
	 \label{table:segperf}
\end{table}

\begin{table}[H]
\setlength{\tabcolsep}{2pt}
\renewcommand{\arraystretch}{1.2}
	 	\centering
	 \begin{tabular}{ |c|c|c|c|}
	 	
	 	\hline
	 	\textbf{Inhomogeneity} & \textbf{Methods}  & \textbf{Mixture Inference}  & \textbf{SO$_2$ inference } \\
         & \textbf{compared} & \textbf{improvement} & \textbf{improvement} \\
	 	\hline
        Level 1 & \textbf{APRECOT vs UP} & 0.84 (0.71 - 0.90) & 0.50 (0.30 - 0.76) \\
        & \textbf{APRECOT vs FCLU} & 0.91 (0.79 - 0.96)  & 0.87 (0.78 - 0.94) \\
        \hline
        Level 2 & \textbf{APRECOT vs UP} & 0.77 (0.65 - 0.86)  & 0.38 (0.26 - 0.71)  \\
        & \textbf{APRECOT vs FCLU} & 0.88 (0.74 - 0.94) & 0.85 (0.78 - 0.91) \\
        \hline
        Level 3 & \textbf{APRECOT vs UP} & 0.72 (0.57 - 0.81) & 0.39 (0.19 - 0.69) \\
        & \textbf{APRECOT vs FCLU} & 0.85 (0.68 - 0.92)  & 0.84 (0.66 - 0.91) \\
	 	\hline
	 	
	 \end{tabular}
	 \caption{Improvement of chromophore mixtures and oxygen saturation (SO$_2$) prediction accuracy of APRECOT compared to UP and FCLU expressed as median relative improvement and interquartile range. vs - versus}
	 \label{table:inhomoglev}
\end{table}
\end{document}